\documentclass[twocolumn,10pt]{IEEEtran}
\renewcommand{\baselinestretch}{0.99}
\usepackage{latexsym}
\usepackage{amsmath, graphics,amssymb,epsfig,subfigure}
\usepackage{cite}
\usepackage{url}
\usepackage{color,soul}
\usepackage{hhline}
\usepackage{bbold}
\usepackage{algorithm}
\usepackage{algorithmic}
\usepackage{graphicx}
\usepackage{bm}

\def\BibTeX{{\rm B\kern-.05em{\sc i\kern-.025em b}\kern-.08em
    T\kern-.1667em\lower.7ex\hbox{E}\kern-.125emX}}

\newfont{\bb}{msbm10 scaled 1000}

\newcommand{\be}{\begin{equation}}
\newcommand{\ee}{\end{equation}}
\newcommand{\bea}{\begin{eqnarray}}
\newcommand{\eea}{\end{eqnarray}}
\newcommand{\bitem}{\begin{itemize}}
\newcommand{\eitem}{\end{itemize}}

\makeatletter
\def\hlinewd#1{%
\noalign{\ifnum0=`}\fi\hrule \@height #1 \futurelet \reserved@a\@xhline}

\makeatother

\makeatletter

\newcommand*{\rom}[1]{\expandafter\@slowromancap\romannumeral #1@}
\makeatother

\newcommand{\tabincell}[2]{\begin{tabular}{@{}#1@{}}#2\end{tabular}}
\usepackage{multirow}
\usepackage{diagbox}
\usepackage{makecell}

\begin{document}

\title{\LARGE{A Codebook Design for FD-MIMO Systems with Multi-Panel Array}}
\vspace{0mm}

\author{\IEEEauthorblockN{Zhilin Fu, Sangwon Hwang, Jihwan Moon, \textit{Member}, \textit{IEEE}, Haibao Ren, and Inkyu Lee, \textit{Fellow}, \textit{IEEE} \\ \vspace*{-6mm}}

\thanks{\copyright 2022 IEEE. Personal use of this material is permitted. Permission from IEEE must be obtained for all other uses, in any current or future media, including reprinting/republishing this material for advertising or promotional purposes, creating new collective works, for resale or redistribution to servers or lists, or reuse of any copyrighted component of this work in other works.}

\thanks{Z. Fu, S. Hwang and I. Lee are with the School of
Electrical Engineering, Korea University, Seoul 02841, South Korea (e-mail:
celynnfu@korea.ac.kr; tkddnjs3510@korea.ac.kr; inkyu@korea.ac.kr).}% <-this % stops a space

\thanks{J. Moon is with the Department of Mobile Convergence Engineering, Hanbat National University, Daejeon 34158, South Korea (e-mail: anschino@staff.hanbat.ac.kr).}% <-this % stops a space

\thanks{H. Ren is with the Wireless Network Research Department, Huawei
Technologies, Shanghai 201206, China (e-mail: renhaibao@huawei.com).}% <-this % stops a space

% \thanks{Digital Object Identifier ……}

}

% \markboth{Journal of \LaTeX\ Class Files,~Vol.~x, No.~x, August~2022}%

% \IEEEpubid{0000--0000/00\$00.00~\copyright 2015 IEEE}

\maketitle

%%%%%%%%%%%%%%%%%%%%%%%%%%%%%%%%%%%%%%%%% Roman Number
\newcommand{\RNum}[1]{\uppercase\expandafter{\romannumeral #1\relax}}
%%%%%%%%%%%%%%%%%%%%%%%%%%%%%%%%%%%%%%%%%%%%%%%%%%%%%%%%%%%%%%%%%%%%%%%%%%%%%%%%%%%%%%%%%%%%%%%%%%%%%%%ABStract
\begin{abstract}
In this work, we study codebook designs for full-dimension multiple-input multiple-output (FD-MIMO) systems with a multi-panel array (MPA). We propose novel codebooks which allow precise beam structures for MPA FD-MIMO systems by investigating the physical properties and alignments of the panels. We specifically exploit the characteristic that a group of antennas in a vertical direction exhibit more correlation than those in a horizontal direction. 
This enables an economical use of feedback bits while constructing finer beams compared to conventional codebooks. The codebook is further improved by dynamically allocating the feedback bits on multiple parts such as beam amplitude and co-phasing coefficients using reinforcement learning.
The numerical results confirm the effectiveness of the proposed approach in terms of both performance and computational complexity.
\vspace*{-1mm}
\end{abstract}

%\begin{IEEEkeywords}
%UAV communication, trajectory optimization, throughput maximization, successive convex approximation
%\end{IEEEkeywords}
\section{Introduction} \label{sec:introduction}

\IEEEpubidadjcol

Recently, demands for real-time connectivity and insufficient bandwidth usage have become critical issues in the telecommunications field. In response to these needs, a multiple input multiple output (MIMO) technology have been actively studied as a key feature for the fifth generation (5G) system \cite{LLu:14}, since they provide increased throughput with extra degrees of freedom and high-speed connections \cite{JCN:2022}. However, its large linear antenna array at a base station (BS) within a limited space generally makes the operational work cumbersome \cite{YHuang:18}. 

Long-Term Evolution (LTE) by Third Generation Partnership Project (3GPP) has included a full-dimension MIMO (FD-MIMO) architecture where antenna elements are positioned in a uniform planar array (UPA) \cite{QNadeem:19} to reduce the required space. Furthermore, a multi-panel based FD-MIMO has been proposed in \cite{3GPP38.214:19}, which assembles antenna elements on different panels. Such a multi-panel array (MPA) has raised a lot of interest for millimeter-wave (mmWave) MIMO systems due to the cost reduction and power savings \cite{YHuang:18}.

The MPA systems can also provide high-resolution channel state information (CSI) and allow an efficient beam management with a higher degree of spatial freedom \cite{ahmadi20195g}, which result in beamforming gains and broader coverage. 
One may consider adopting conventional mmWave MIMO codebooks \cite{R1_ZXiao16, R1_ZXiao17, R1_StatisticallyCB} to the MPA in a similar manner. The authors in \cite{R1_StatisticallyCB} presented a codebook that minimizes the average distortion with a fully digital precoding architecture. For hybrid precoding which composes both analog and digital precoders, hierarchical codeword search schemes were proposed in \cite{R1_ZXiao16}, \cite{R1_ZXiao17} and \cite{R1_CQi20}. In these works, based on discrete Fourier transform (DFT) codebooks, the mmWave subarray characteristics were utilized to obtain the optimal beam direction. Besides, some research has been done which exploits the correlation between codebooks and channel spatial characteristics \cite{R1_DYang10}\cite{R1_JLi13}.
However, in the MPA systems, each panel carries its own radio frequency (RF) circuits, and thus incurs a phase offset problem during calibrating RF chains in different panels, even if the panels are closely spaced \cite{R1-1702071:17}. As a result, existing single-panel codebook designs based on Kronecker product and discrete Fourier transform (DFT) \cite{JSong:17} cannot be directly applied to the MPA systems.

\IEEEpubidadjcol

Recently, a multi-panel codebook, which is called the Type-\RNum{1} multi-panel (MP) codebook \cite{3GPP38.214:19}, was introduced which quantifies channel direction information (CDI) and generates a DFT codebook of the entire MPA. The integral DFT codebook is then calibrated by co-phasing factors in the panel. Also, the authors of \cite{HMShin:18} proposed the independent panel codebook (IPC), which is a modified Type-\RNum{1} MP codebook by independently quantifying each panel's CDI to address the phase ambiguity (PA) among panels.

Beside Type-\RNum{1} codebook, another codebook called Type-\RNum{2} was presented in \cite{R1-1709232:17}, which has a higher precision of a beamforming structure by simultaneously adjusting the amplitudes and power of the over-sampled DFT beams. This Type-\RNum{2} scheme can potentially enhance the performance of the MPA systems, but to the best of the authors' knowledge, the Type-\RNum{2} based MPA codebooks have not been fully optimized yet. One of the main reasons is that an uncareful adoption of Type-\RNum{2} onto the MPA may result in an overwhelmingly large number of feedback bits.

In this work, we propose novel Type-\RNum{2} codebooks for MPA systems by investigating the physical properties and alignments of the panels to reduce the feedback burden. Noting that quantizing each panel may require too many feedback bits, we group some panels as a single one in a way that an accurate directivity property of Type-\RNum{2} can be preserved as much as possible. We particularly utilize the fact that multiple antennas stacked in the vertical direction create a radiation pattern that is thinner and wider \cite{balanis2016antenna}. Hence, in our scheme, panels in the vertical direction are treated as a single panel and only the PA among these groups is adjusted.

% \vspace{-1mm}
The proposed codebook is composed of different components such as beam amplitude, beam combining coefficients, DFT size, and co-phasing coefficients. To further improve the codebook performance, we also introduce a deep Q-learning (DQL) based reinforcement learning (RL) algorithm to identify the best bit allocation for each component. The numerical results verify that our proposed schemes exhibit an outstanding performance-complexity trade-off compared with conventional schemes.

\vspace{-2mm}
\section{System Model} 
\label{sec:system_model and problem formulation}
% \vspace{-1mm}

\begin{figure}
\begin{center}
\includegraphics[width=2.5in]{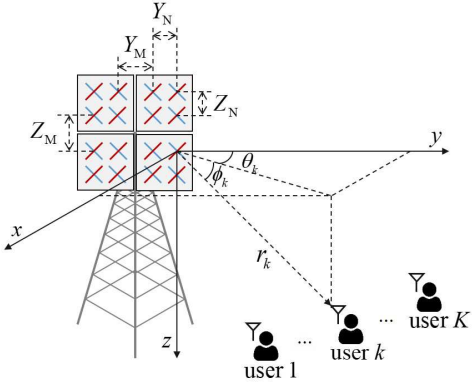}  % 2column
\end{center}
\vspace{-2mm}
\caption{Multi-user FD-MIMO systems with MPA}
\label{fig:system_model}
\vspace*{-4mm}
\end{figure}

As illustrated in Fig. \ref{fig:system_model}, we consider an MPA system where a BS is equipped $M_\mathrm{v}$ vertical panels and $M_\mathrm{h}$ horizontal panels, each of which contains a uniformly spaced rectangular antenna array (URA) with $N_\mathrm{v}$ and $N_\mathrm{h}$ antennas in vertical and horizontal direction, respectively. At the receiving end, $K$ single-antenna users are served with the same frequency/time resources from the BS.

\vspace{-3mm}
\subsection{Spatial Channel Model for Multi-Panel MIMO Systems}
% \vspace{-1mm}

We adopt the spatial channel model (SCM) proposed by the 3GPP standards \cite{QNadeem:19}\cite{WLiu:17} that characterizes three dimensional (3D) correlation among antenna elements. Particularly, the authors in \cite{HMShin:18} have studied the 3D correlation model with different spacing in horizontal and vertical directions. Assuming a multi-path channel with  single polarized antenna elements, the aggregate channel matrix $\mathbf{H}_{k} \in \mathbb{C}^{M_{\mathrm{v}} N_{\mathrm{v}} \times M_{\mathrm{h}} N_{\mathrm{h}}}$ between the BS and the $k$-th user is given by
\vspace{-1mm}
\begin{equation}
    \mathbf{H}_{k}=\frac{1}{\sqrt{S}} \sum_{i=1}^{S} \mathbf{H}_{k}^{i} , 
    \label{H_mat}
    \vspace{-1mm}
\end{equation}
where $S$ represents the number of paths, and $\mathbf{H}_{k}^{i}$ stands for the channel matrix along the $i$-th path. 

Let us define $\rho_{k}^{i} = z_{i} 
{10^{ {- \gamma\left(r_{k}\right)} / {20} }
}$ as the large-scale fading coefficient for the $i$-th path where $z_{i}$ is the random complex coefficient with zero mean and unit variance, and $\gamma\left(r_{k}\right)$ indicates the path-loss with the distance $r_{k}$ between the BS and the user.
In the SCM, the phase of channel elements varies by the panel arrays, while the magnitude is fixed regardless of the panel arrays. For the $i$-th path and the $k$-th user, denote $\phi_{k}^{i}$ and $\theta_{k}^{i}$ as the azimuth angle of departure (AOD) and the zenith angle of departure (ZOD), respectively. 
Here, we designate $\left[ \mathbf{A} \right]_{m,n}$ as the $\left(m,n\right)$-th element of a matrix $\mathbf{A}$. For the MPA systems, the $\left(m, n\right)$-th element of the channel matrix ($m= 1,2, \dots, M_{\mathrm{v}} N_{\mathrm{v}}$, $n= 1,2, \dots, M_{\mathrm{h}} N_{\mathrm{h}}$) along the $i$-th path $\mathbf{H}_{k}^{i}$ is expressed as
\begin{equation}
    % \textstyle
    [ \mathbf{H}_{k}^{i} ]_{m, n} \!\! = \! 
    \rho_{k}^{i} \!
    \exp \! \left( - \! j \! { \frac{2 \pi}{\lambda}} \! \left(  \Delta_{\mathrm{v}}^{m} \! \sin \phi_{k}^{i} \! + \! \Delta_{\mathrm{h}}^{n} \! \cos \theta_{k}^{i} \! \cos \phi_{k}^{i}  \right) \! \right) \! ,
    \label{array_response_SPol}
    \vspace{-1mm}
\end{equation}
where the $\lambda$ means the wavelength, and $\Delta_{\mathrm{v}}^{m}$ and $\Delta_{\mathrm{h}}^{n}$ specify the spacing of antenna elements for the vertical and horizontal domain, respectively as 
\begin{displaymath}
    \begin{aligned}
        \Delta_{\mathrm{v}}^{m} &= Z_{\mathrm{M}} \left(\left \lceil\tfrac{m}{N_{\mathrm{v}}}\right\rceil - 1 \right) + Z_{\mathrm{N}} \left( m - \left\lceil\tfrac{m}{N_{\mathrm{v}}}\right\rceil \right) , \\
        \Delta_{\mathrm{h}}^{n} &= Y_{\mathrm{M}} \left(\left \lceil\tfrac{n}{N_{\mathrm{h}}}\right\rceil - 1 \right) + Y_{\mathrm{N}} \left( n - \left\lceil\tfrac{n}{N_{\mathrm{h}}}\right\rceil \right).
    \end{aligned}
    \vspace{-1mm}
\end{displaymath}
Here, $Z$ and $Y$ correspond to the spacing values in the vertical and horizontal domain, respectively, $Z_{\mathrm{M}}$ and $Y_{\mathrm{M}}$ indicate the panel spacing distance, $Z_{\mathrm{N}}$ and $Y_{\mathrm{N}}$ equal the distance between antenna elements on the same panel, and $\lceil \cdot \rceil$ denotes the ceiling operation.

With these in hand, the vectorized form of the wireless channel $\mathbf{H}_{k}$ for the $k$-th user can be introduced by utilizing the vectorization operation vec($\cdot$) as
\begin{equation}
    \mathbf{h}_{k}=\frac{1}{\sqrt{M_{\mathrm{v}} N_{\mathrm{v}} M_{\mathrm{h}} N_{\mathrm{h}}}} \operatorname{vec}\left(\mathbf{H}_{k}\right) ,
    \label{vectorization}
    \vspace{-1mm}
\end{equation}
in which $\mathbf{h}_{k}$ becomes a complex column vector of length $ M_{\mathrm{v}} N_{\mathrm{v}} M_{\mathrm{h}} N_{\mathrm{h}}$.

\vspace{3mm}
\subsection{1-to-M Mapping}
% \vspace{-1mm}

Ideally, each antenna on FD-MIMO can be equipped with a dedicated RF chain \cite{JLee20}. This approach, however, requires high cost in terms of hardware and energy. To address this issue, separate analog and digital beamforming called 1-to-$M$ mapping \cite{QNadeem:19} can be utilized in which $M$ physical antenna elements are combined into one RF chain to reduce the complexity of digital beamforming, while additional analog beam adjustments are carried out by phase shifters.

For each RF chain, $R_{\mathrm{v}}$ vertical antenna elements are assembled through the 1-to-$M$ mapping approach with $M = R_{\mathrm{v}}$.
The $m$-th element of the weight vector $\mathbf{u} \in \mathbb{C}^{R_{\mathrm{v}} \times 1}$ with these $R_{\mathrm{v}}$ antenna elements is obtained as
\begin{equation}
    % \textstyle
    [ \mathbf{u} ]_{m} = 
    \frac{1}{\sqrt{ R_{\mathrm{v}} }} \! \exp \! \left( -j\frac{2\pi}{\lambda} \! \left( m \! - \! 1\right) \! \Delta_{\mathrm{v}}^{m} \cos \theta_{{\mathrm{tilt}}} \! \right) \! ,
    \label{TXRU}
    \vspace{-1mm}
\end{equation}
where $\theta_{tilt}$ indicates the steering angle in the vertical direction.

The number of RF chains on each panel equals $N_{\mathrm{v}}^{\mathrm{R}} \times N_{\mathrm{h}}$, where $N_{\mathrm{v}}^{\mathrm{R}} = N_{\mathrm{v}} / R_{\mathrm{v}}$. Then, by applying the analog beamforming weight component to the 3D channel coefficients for each antenna element, the $\left( m, n \right)$-th elements ($m= 1,2, \dots, M_{\mathrm{v}} N_{\mathrm{v}}^{\mathrm{R}}$, and $n= 1,2, \dots, M_{\mathrm{h}} N_{\mathrm{h}}$) can be represented as
\begin{equation}
    [ \widetilde{ \mathbf{H} }_{k} ]_{m, n}=
    \mathbf{u}^{H} \check{\mathbf{h}}_{k},
    \vspace{-1mm}
\end{equation}
in which $\check{\mathbf{h}}_{k} \in \mathbb{C}^{ R_{\mathrm{v}} \times 1 }$ is the channel coefficient corresponding to the antenna elements with the same RF chain. By operating the vectorization on $\widetilde{ \mathbf{H} }_{k}$, the corresponding channel vector for the $k$-th user can be denoted as $\widetilde{ \mathbf{h} }_{k} \in \mathbb{C}^{ M_{\mathrm{v}} N_{\mathrm{v}}^{\mathrm{R}} M_{\mathrm{h}} N_{\mathrm{h}} \times 1} $ similar to \eqref{vectorization}.

\vspace{-1mm}
\section{Type-\RNum{2} single panel codebook}
\label{sec:Type2}
% \vspace{-1mm}

In this section, we first review Type-\RNum{2} single panel (SP) codebook systems \cite{3GPP38.214:19}, where the CSI feedback framework consists of two matrices selected from separate codebooks. To apply this SP codebook design to our system model, we assume the MPA as a whole panel. The total number of vertical antenna ports and horizontal antenna ports are then expressed as $N_{\mathrm{v}}^{\mathrm{t}} = M_{\mathrm{v}} N_{\mathrm{v}}^{\mathrm{R}}$ and $N_{\mathrm{h}}^{\mathrm{t}} = M_{\mathrm{h}} N_{\mathrm{h}}$, respectively. For further consideration with the cross-polarization antenna model, the total number of antenna ports should be doubled as $2 N_{\mathrm{v}}^{\mathrm{t}} N_{\mathrm{h}}^{\mathrm{t}}$. Defining $L$ and $R$ as the number of over-sampled 2D DFT beams and the rank of the codebook, respectively, 
the basic form of the precoding matrix index (PMI) codebook is given by 
\vspace{-1mm}
\begin{equation}
    \mathbf{W} =\mathbf{W}_{1} \mathbf{W}_{2} ,
    \label{PMI_mat_W}
    \vspace{-1mm}
\end{equation}
where $\mathbf{W}_{1} \in \mathbb{C}^{ {2 N_{\mathrm{v}}^{\mathrm{t}} N_{\mathrm{h}}^{\mathrm{t}} } \times 2L}$ represents the wide-band feedback which provides a group of beams based on long-term channel statistic, and $\mathbf{W}_{2} \in \mathbb{C}^{2L\times R}$ denotes the relatively short-term and wide-band/sub-band (WB/SB) feedback \cite{asplund2020advanced}.

According to this mechanism, the Type-\RNum{2} SP codebook based CSI feedback can report both the WB and the SB amplitude information of the selected beams. To support a cross-polarized antenna array as adopted in 3GPP standard, the beam selection matrix $\mathbf{W}_{1}$ is composed of unconstrained beam selection from an orthogonal basis\cite{R1-1709232:17} as 
\vspace{-1mm}
\begin{equation}
    \mathbf{W}_{1} = \mathrm{diag} \left( \mathbf{B},~ \mathbf{B} \right) ,
    \label{PMI_mat_W1}
    \vspace{-1mm}
\end{equation}
where $\mathrm{diag} \left( \cdot \right)$ defines a block diagonal matrix containing each matrix on the diagonal and $\mathbf{B} \in \mathbb{C}^{ {N_{\mathrm{v}}^{\mathrm{t}} N_{\mathrm{h}}^{\mathrm{t}} } \times L}$ accounts for the antenna subgroup with the same polarization.

Since the correlation is high within the same antenna subgroup, it makes sense to use a grid of beam codebooks implemented from the DFT based precoder vectors \cite{QNadeem:19}, and thus $\mathbf{B}$ is constructed by adjacent DFT vectors. 
To be specific, in the Type-\RNum{2} SP codebook design, the DFT codebook $\mathbf{B}$ consists of $L \in \{2,3,4\}$ over-sampled 2D DFT beams with length $N_{\mathrm{v}}^{\mathrm{t}} N_{\mathrm{h}}^{\mathrm{t}}$. Each over-sampled beam can be denoted as $\mathbf{b}_{\theta_{\mathrm{v}}^{\left( i \right)}, \theta_{\mathrm{h}}^{\left( i \right)}}$ for $i = 1, \dots , L$, where 
\vspace{-1mm}
\begin{displaymath}
    \theta_{\mathrm{v}}^{\left( i \right)} = 2^{B_{\mathrm{v}}} n_{\mathrm{v}}^{\left( i \right)} + q_{\mathrm{v}} ,~ \text{and}~
    \theta_{\mathrm{h}}^{\left( i \right)} = 2^{B_{\mathrm{h}}} n_{\mathrm{h}}^{\left( i \right)} + q_{\mathrm{h}}
    \vspace{-1mm}
\end{displaymath}
refer to the indices for the $i$-th over-sampled beam vector in vertical and horizontal domain, respectively. Here, $B_{\mathrm{v}}$ and $B_{\mathrm{h}}$ are the number of feedback bits corresponding to the selected orthogonal beams, and $q_{\mathrm{v}} \in \{0,~...,~ 2^{B_{\mathrm{v}}}-1\}$ and $q_{\mathrm{h}} \in \{0,~...,~ 2^{B_{\mathrm{h}}}-1\}$ represent the beam rotation factors. In order to make the over-sampled beams orthogonal, the beam indices $n_{\mathrm{v}}^{(i)} \in \{0,~...,~ N_{\mathrm{v}}^{\mathrm{t}}-1\}$ and $n_{\mathrm{h}}^{(i)} \in \{0,~...,~ N_{\mathrm{h}}^{\mathrm{t}}-1\}$ should be $( n_{\mathrm{v}}^{( i)},~ n_{\mathrm{h}}^{(i)}) \neq (n_{\mathrm{v}}^{(j)},~ n_{\mathrm{h}}^{(j)})$ for $i\neq j$.
Consequently, the required number of bits for feedback information on the beam selection is 
\vspace{-1mm}
\begin{equation}
    B_{\mathrm{DFT}} = B_{\mathrm{v}} + B_{\mathrm{h}} + \left\lceil\log_{2} \tbinom{ N }{L} \right\rceil .
    \label{B_DFT}
    \vspace{-1mm}
\end{equation}
where the $\left\lceil\log_{2} \tbinom{N}{L} \right\rceil$ bits determine which $L$ beams are over-sampled among $N = N_{\mathrm{v}}^{\mathrm{t}} N_{\mathrm{h}}^{\mathrm{t}}$ possible beams.

We now consider the short-term feedback matrix $\mathbf{W}_{2}$ which performs a weighted combination of $L$ beams and co-phase adjustment between polarization. We denote $c_{r,l,i}$ as the beam combining coefficient and $p_{r,l,i}^{\left(\mathrm{WB/SB}\right)}$ as the WB/SB beam amplitude scaling factor for beam $i$ for polarization $r$ and layer $l$. 
Without loss of generality, we adopt the rank one scheme ($R=1$) with the WB-only amplitude scaling mode of the Type-\RNum{2} SP codebook \cite{R1-1709232:17} for simplicity. This means that $\mathbf{W}_{2}$ has only one layer ($l=1$) and $p_{r, i}^{(\mathrm{SB})} \! = \! 1$.

Assuming that the feedback information of the beam combining coefficient $c_{r,i}$ and the beam amplitude scaling factor $p_{r,i}^{\left(\mathrm{WB}\right)}$ are respectively expressed with $B_{\mathrm{c}}$ and $B_{\mathrm{p}}$ feedback bits, $\mathbf{W}_{2}$ can be formed as
\begin{equation}
    \mathbf{W}_{2}=\left[
    1,
    c_{1,2}  p_{1,2}^{(\mathrm{WB})}
    \!,\dots, \!
    c_{r,i}  p_{r,i}^{(\mathrm{WB})}
    \!,\dots, \!
    c_{2,L}  p_{2,L}^{(\mathrm{WB})}
    \right]^T , 
    \label{PMI+mat_W2}
\end{equation}
where $c_{r,i} \! \in \! \{ \exp \left( j\frac{n\pi}{2} \right), n=0,1, \cdots, 2^{B_{\mathrm{c}}}-1 \}$ and $p_{r, i}^{(\mathrm{WB})} \! \in \! \{\sqrt{2^{0}}, \sqrt{2^{-1}}, \! \cdots, \! \sqrt{2^{-\left(B_{\mathrm{p}}-2\right)}}, 0\}$. For the over-sampled beams with different polarization, $c_{r, i}$ and $p_{r, i}^{(\mathrm{WB})}$ are independent.

Note that one of the over-sampled beams is determined with initial values ($c_{1,1}  p_{1,1}^{(\mathrm{WB})}=1$), so that $\left(2 L-1\right)\left(B_{\mathrm{p}}+B_{\mathrm{c}}\right)$ bits are required for the short-term feedback information.
Hence, the total required number of feedback bits $B_{\mathrm{\RNum{2}}}$ in cross-polarization antenna systems is computed as 
\begin{equation}
    \begin{aligned}
    B_{\mathrm{\RNum{2}}}=& B_{\mathrm{DFT}} + \left\lceil\log _{2}(2 L)\right\rceil+\left(2 L-1\right)\left(B_{\mathrm{p}}+B_{\mathrm{c}}\right),
    \end{aligned}
    \label{Bit_type2}
\end{equation}
where $\left\lceil\log _{2}(2 L)\right\rceil$ bits are used to label the over-sampled $2 L$ beams. The codebook length which accounts for the search complexity of the channel quantization for the Type-\RNum{2} codebook design is then calculated as
\begin{equation}  % 2column
    {\Omega}_{\mathrm{\RNum{2}}} \!
    = \! {\Omega}_{\mathbf{W}_{\!1}} {\Omega}_{\mathbf{W}_{\!2}}
    = \! 2 L N_{\mathrm{v}}^{\mathrm{t}} N_{\mathrm{h}}^{\mathrm{t}} 2^{B_{\mathrm{v}}+B_{\mathrm{h}}} \times ( 2 L - 1 ) 2^{B_{\mathrm{p}}+B_{\mathrm{c}}}.
    \label{CBLength_type2}
\end{equation}
This indicates that the search complexity grows with the number of antenna ports for typical MPA systems. To address this issue, we propose a new codebook scheme for MPA designs in the next section.

% \vspace{-2mm}
\section{Proposed Type-\RNum{2} Multi-panel Codebook}
\label{sec:proposed}

In this section, we propose a new line-panel (LP) codebook which extends the Type-\RNum{2} SP codebook to the multi-panel case with cross-polarization antennas.
The LP codebook groups panels in the same vertical direction as $M_{\mathrm{LP}}=M_{\mathrm{h}}$ line-panels with $N_{\mathrm{LP}}=M_{\mathrm{v}}N_{\mathrm{v}}^{\mathrm{R}} \times N_{\mathrm{h}}$ antenna ports on each line-panel. For the cross-polarization antenna model, the number of antenna ports on each line-panel becomes $2 N_{\mathrm{LP}}$.

Let us denote 
$\mathbf{\mathcal{C}}_{\mathrm{SLP}} = \left\{ \mathbf{c}_{\mathrm{SLP}}^{\left( 1 \right)}, \mathbf{c}_{\mathrm{SLP}}^{\left( 2 \right)}, \dots, \mathbf{c}_{\mathrm{SLP}}^{\left( {\Omega}_{\mathrm{S}} \right)} \right\}$
as the single line-panel (SLP) codebook consisting of ${\Omega}_{\mathrm{S}}$ candidate precoding vectors $\mathbf{c}_{\mathrm{SLP}}^{\left( i \right)} \in \mathbb{C}^{2 N_{\mathrm{LP}} \times 1}$ for $i=1, \dots, {\Omega}_{\mathrm{S}}$.
In $\mathbf{\mathcal{C}}_{\mathrm{SLP}}$, each SLP codeword candidate $\mathbf{c}_{\mathrm{SLP}}^{\left( i \right)}$ is generated following the Type-\RNum{2} SP codebook scheme as a unique combination of different $L$ DFT beams with diverse co-phasing coefficients, amplitude scaling factors, and cross-polarization. Referring to \eqref{CBLength_type2}, the length of $\mathbf{\mathcal{C}}_{\mathrm{SLP}}$ can be calculated as 
\begin{equation}
    {\Omega}_{\mathrm{S}} = 2 L ( 2 L - 1 ) N_{\mathrm{LP}} 2^{B_{\mathrm{v}} + B_{\mathrm{h}} + B_{\mathrm{p}}+B_{\mathrm{c}}} .
    \label{CBLength_SLP}
    \vspace{-1mm}
\end{equation}

Then, each user quantizes the channel vector corresponding to each line-panel, independently, for $m = 1, 2, \dots, M_{\mathrm{LP}}$ as
\begin{equation}
    \begin{aligned}
    ~\mathbf{\bar{h}}_{m, k} \! = \mathbf{c}_{\mathrm{SLP}}^{\left( i_{m, k}^* \right)} \!,~ \!
    \text{where ~} \!
    i_{m, k}^* \! = \mathop{\arg\max}\limits_{1 \leq i \leq {\Omega}_{\mathrm{S}}} \left| \widetilde{\mathbf{h}}_{m, k}^{H} \mathbf{c}_{\mathrm{SLP}}^{\left( i \right)} \right| ,
    \label{DFT_selection}
    \end{aligned}
    \vspace{-1mm}
\end{equation}
and $\widetilde{\mathbf{h}}_{m, k}$ is formed by a part of elements in $\widetilde{\mathbf{h}}_{k}$ which corresponds to the same line-panel.
As a result, the DFT beam candidate $\mathbf{c}_{\mathrm{SLP}}^{\left( i_{m, k}^* \right)}$ is chosen which generates the maximum inner product with the corresponding channel vector $\widetilde{\mathbf{h}}_{m, k}^{H}$.

Based on the SLP codebook, we now design the LP codebook for multiple panels. The LP codebook adds the panel co-phasing factors, which capture panel-wise channel characteristics. With $B_{\mathrm{LP}}$ feedback bits for the panel co-phasing coefficients, the uniform PA codebook $\mathbf{\mathcal{C}}_{\mathrm{PA}}$ is defined as a $M_{\mathrm{LP}} \times \left( \! M_{\mathrm{LP}} \!-\! 1 \! \right)2^{B_{\mathrm{LP}}}$ matrix whose columns are given by $\left[ 1, e^{j\theta_{2}}, \dots, e^{j\theta_{M_{\mathrm{LP}}}} \right]^{T}$ with ${\theta}_m \in \big\{ 0,\frac{2\pi}{2^{B_{\mathrm{LP}}}},\dots, \frac{2\pi(2^{B_{\mathrm{LP}}}-1)}{2^{B_{\mathrm{LP}}}}\big\}$ for $m = 2, 3, \dots, M_{\mathrm{LP}}$. 
Here, the original phase of the first line-panel is adopted as a reference for the phase adjustment of other panels, and the length of the whole panel co-phasing codebook $\mathbf{\mathcal{C}}_{\mathrm{PA}}$ can be expressed as ${\Omega}_{\mathrm{P}} = \left( M_{\mathrm{LP}}-1 \right) 2^{B_{\mathrm{LP}}}$.

The LP codebook ${\mathbf{\mathcal{C}}}_{\mathrm{LP}}= \left\{ \mathbf{c}_{\mathrm{LP}}^{\left( 1 \right)}, \mathbf{c}_{\mathrm{LP}}^{\left( 2 \right)}, \dots, \mathbf{c}_{\mathrm{LP}}^{\left( {\Omega}_{\mathrm{P}} \right)} \right\}$ can be obtained by performing dot products between the aggregation of quantized SLP codewords 
$\left[\mathbf{\bar{h}}_{1, k}^T, \mathbf{\bar{h}}_{2, k}^T, \dots, \mathbf{\bar{h}}_{M_{\mathrm{LP}}, k}^T \right]^T$ and each column of $\mathbf{\mathcal{C}}_{\mathrm{PA}}$, where $\bm{\mathbf{c}}_{\mathrm{LP}}^{\left( i \right)} \in \mathbb{C}^{M_{\mathrm{LP}} N_{\mathrm{LP}} \times 1}$ for $i=1, \dots, {\Omega}_{\mathrm{P}}$ is given by
\begin{equation}
    \mathbf{\mathbf{c}}_{\mathrm{LP}}^{\left( i \right)} = {\left[ 
    \mathbf{\bar{h}}_{1, k}^T
    ,
    e^{j\theta_{2}}\mathbf{\bar{h}}_{2, k}^T
    , \dots,
    e^{j\theta_{M_{\mathrm{LP}}}}\mathbf{\bar{h}}_{{M_{\mathrm{LP}}}, k}^T
    \right]}^{T}.
    \vspace{-1mm}
    \label{LPC}
\end{equation}
Then, each user separately quantizes its aggregate channel vector $\widetilde{\mathbf{h}}_{k} \in \mathbb{C}^{M_{\mathrm{LP}}N_{\mathrm{LP}} \times 1}$ and obtains the suitable panel co-phasing coefficients from ${\mathbf{\mathcal{C}}}_{\mathrm{LP}}$ as
\begin{equation}
    \begin{aligned}
    ~ \widehat{\mathbf{h}}_k =& \mathbf{c}_{\mathrm{LP}}^{\left( i_{k}^* \right)} \! ,~
    \text{where~}
    i_{k}^* \! = \mathop{\arg\max}\limits_{1 \leq i \leq {\Omega}_{\mathrm{P}}} \left| \widetilde{\mathbf{h}}_{k}^{\mathrm{H}} \mathbf{c}_{\mathrm{LP}}^{\left( i \right)} \right| . 
    \label{LPC_selection}
    \end{aligned}
    \vspace{-1mm}
\end{equation}
The total required number of feedback bits $B_{\mathrm{L}}$ in this proposed scheme is then calculated as
\begin{equation}
    \begin{aligned}
    B_{\mathrm{L}}=& M_{\mathrm{LP}} \big( B_{\mathrm{DFT}} + \left\lceil\log_{2}(2 L)\right\rceil \\
    &+\left(2 L-1\right)\left(B_{\mathrm{p}}+B_{\mathrm{c}}\right) \big) + \left(M_{\mathrm{LP}}-1\right)B_{\mathrm{LP}},
    \end{aligned}
    % \vspace{-1mm}
    \label{Bit_LPC}
\end{equation}
where $B_{\mathrm{DFT}}$ can be calculated by \eqref{B_DFT} with $ N = N_{\mathrm{LP}}$.
Note that if the Type-\RNum{2} SP codebook in \eqref{Bit_type2} is directly applied to MPA by merely multiplying the required number of bits by panels, the total number of bits becomes multiples of $B_{\mathrm{L}}$ in \eqref{Bit_LPC}. It is clear that our proposed LP codebook design requires much less bits.

Compared with the Type-\RNum{2} SP scheme that neglects panel interspace, the LP codebook scheme can mitigate the PA issue with only a small amount of extra feedback bits.  Furthermore, in the LP codebook scheme, the channel vectors corresponding to each line-panel are quantized individually. In contrast, the Type-\RNum{2} SP codebook scheme quantizes the channel vectors corresponding to the whole MPA at once. 
As a result, for the LP codebook, the size of the codeword candidate matrices is much smaller, and the resulting search complexity is substantially lower as
\begin{equation}
    \begin{aligned}
    {\Omega}_{\mathrm{L}} \!
    &=\!M_{\mathrm{LP}} {\Omega}_{\mathrm{S}} + {\Omega}_{\mathrm{P}} \\
    &=\! 2 L ( 2 L - 1 ) M_{\mathrm{LP}} N_{\mathrm{LP}} 2^{B_{\mathrm{v}} \!+\! B_{\mathrm{h}} \!+\! B_{\mathrm{p}} \!+\! B_{\mathrm{c}}}  \!+\! \left( M_{\mathrm{LP}} \!-\! 1 \right) 2^{B_{\mathrm{LP}} }.
    \end{aligned}\label{CBLength_LPC}
    % \vspace{-1mm}
\end{equation}
In the meantime, assuming that the total number of feedback bits is
$B$, the search complexity of the conventional 2D DFT codebook is
% \vspace{-1mm}
\begin{equation}
    {\Omega}_{\mathrm{DFT}}=2^B .
    \label{CBLength_DFT}
    \vspace{-1mm}
\end{equation}
The advantage of the LP codebook is demonstrated in Section \RNum{6} which compares ${\Omega}_{\mathrm{DFT}}$, ${\Omega}_{\mathrm{\RNum{2}}}$, and ${\Omega}_{\mathrm{L}}$.

% \vspace{-1mm}
\section{Reinforcement learning based Dynamic Bit Allocation}
\label{sec:RL-LPC-DynamicBitAlloc}
% \vspace{-1mm}

The 3GPP has given guidelines for the payload calculation for the Type-\RNum{2} SP codebook with common long-term evolution (LTE) MIMO configurations, with $B_{\mathrm{p}}=3$ and $B_{\mathrm{c}}=2$ \cite{R1-1709232:17}. One feasible approach for the bit allocation issue is to directly adopt the 3GPP guideline to the LP codebook with $B_{\mathrm{LP}}=2$ while allocating the rest of the feedback bits on $B_{\mathrm{DFT}}$. However, it is apparent that such an approach does not guarantee the optimal performance. The complexity of exhaustive search for finding the best bit allocation can also become prohibitive even with a practical number of limited feedback bits. Therefore, in this section, we propose the RL based dynamic bit allocation algorithm to further optimize the codebook.

From the expression of \eqref{Bit_LPC}, one can see that the bit allocations for $B_{\mathrm{LP}}$, $B_{\mathrm{v}}$, $B_{\mathrm{h}}$, $B_{\mathrm{p}}$, and $B_{\mathrm{c}}$ need to be determined. In order to identify these bits, we adopt an DQL based RL algorithm. To this end, the Markov decision process (MDP) can be designed as follows. 
% \cite{mnih2015human}. \
First, at the $n$-th time step, the state $s_{n}$ can be defined as an aggregation of the bit components in \eqref{Bit_LPC} as
% \vspace{-1mm}
\begin{equation}
    s_n = \left\{B_{\mathrm{LP}},~B_{\mathrm{v}},~B_{\mathrm{h}},~B_{\mathrm{p}},~B_{\mathrm{c}}\right\}.
    \label{RL_state}
    \vspace{-1mm}
\end{equation}
Each component in $s_n$ is dynamically selected according to the action $a_{n}$ from the action space $\mathcal{A}$ as
% \vspace{-1mm}
\begin{equation}
    \mathcal{A}=\left\{B_{\mathrm{LP}}^{+},~ B_{\mathrm{LP}}^{-},~ ...,~ B_{\mathrm{p}}^{+},~ B_{\mathrm{p}}^{-},~ B^0\right\},
    \vspace{-1mm}
\end{equation}
where ${(\cdot)}^{\pm}$ means plus/minus one bit on one component, and $B^0$ indicates retaining the same state. 
After acting $a_n$, the remaining bits will be allocated to $B_{\mathrm{c}}$ from (\ref{Bit_LPC}). Then, the state $s_n$ is updated to $s_{n+1}$ while the reward $r_{n}$ is received. Notice that the transition to the state $s_{n+1}$ is only dependent on $s_n$ and $a_{n}$, satisfying the memoryless characteristic of the Markov property.

Next, we represent $\mathnormal{G}_{n+1}$ as the average sum-rate value corresponding to the codebook generated through $s_{n+1}$, and denote $\bar{G}$ as the average sum-rate value simulated from conventional settings. We also define $G_{\mathrm{max}}$ as the maximum average sum-rate value. Then, we utilize the reward $r_{n}$ as
\begin{equation}
    r_{n} \!= \!
      \begin{cases}
        \! \mathrm{\eta} b_{n+1} \left( 1 + 2^{ G_{n+1} - \bar{G}} \right)\!\!\!\! &, \text{if $G_{n+1} = G_{\mathrm{max}}$},\\
        \eta b_{n+1} 2^{ G_{n+1} - \bar{G} }\!\!\!\! &, \text{if $\bar{G} \!\leq\! G_{n+1}$\textless$G_{\mathrm{max}}$},\\
        \eta \log_{2}\left( b_{n+1} \frac{G_{n+1}}{\bar{G}} \right)\!\!\!\! &, \text{otherwise},\\
      \end{cases}
    \label{RL:reward}
\end{equation}
where $\eta$ indicates a constant to emphasize the performance difference among bit allocations, $b_{n+1} = B_{n+1}/{B_n}$ and $B_n$ means the total number of feedback bits at the $n$-th step. Through the reward $r_{n}$, $s_{n+1}$ can be evaluated whether it is an optimal bit allocation scheme or not.

Based on the MDP design above, we propose the DQL based dynamic bit allocation method. In the proposed DQL based method, an agent efficiently determines a bit allocation strategy. Specifically, at each time step, the deep Q-network (DQN) with a trainable parameter $\theta$ approximates the Q-value $Q(s_{n},a_{n};\theta)$ which represents the expected cumulative reward \cite{mnih2015human}. 
The action $a_{n}$ is obtained and conducted based on the $\epsilon$-greedy algorithm. Then, the tuple $\left(s_{n}, a_{n}, r_{n}, s_{n+1}\right)$ is stored into the experience-replay memory $\mathcal{D}$ with length $l_{\mathcal{D}}$.

With mini-batch samples $(s,a,r,s')\!\in\!\mathcal{B}$, the DQN's parameter $\theta$ is updated by the gradient descent method $\theta\leftarrow\theta - \alpha\nabla_{\theta}\mathcal{L}$, where $\alpha$ and $\nabla_{\theta}\mathcal{L}$ represent the learning rate and the gradient of the loss value $\mathcal{L}$, respectively. The loss function $\mathcal{L}$ is formulated as
\begin{equation}
    \begin{aligned}
        \mathcal{L} = \frac{1}{|\mathcal{B}|} \sum_{\mathcal{B}} \left[ r + \gamma \mathop{\max}\limits_{a'} Q\left(s', a'; \theta^- \right) -  Q\left(s, a; \theta\right) \right]^{2},
    \end{aligned}
    \label{RL:Loss}
    % \vspace*{-2mm}
\end{equation}
where $\gamma$ is the discount factor and $Q\left(s', a'; \theta^{-}\right)$ stands for the estimated Q-value of the target Q-network with parameter $\theta^{-}$ \cite{SHwang20}. The detailed procedure of the proposed dynamic bit allocation scheme is summarized in Algorithm \ref{RL:algorithm}.
One can note that our proposed bit allocation algorithm is flexible enough to find the best bit allocation method according to different panel configurations and various $B$. Moreover, our proposed LP codebook is applicable even when $B$ is smaller than the minimum requirement of the 3GPP-based bit allocation method.

\begin{algorithm}
\caption{Proposed Bit Allocation Algorithm}
\label{RL:algorithm}
\begin{algorithmic}
    % \FOR{$e$ = 1 to $E$}
    \STATE Initialize the parameter $\theta$ randomly, and set the target Q-network as $\theta^-=\theta$
    \REPEAT
        \STATE Select action $a_n$ and calculate $s_{n+1}$ by \eqref{Bit_LPC}
        \STATE Calculate the $G_{n\!+\!1}$ and observe $r_{\!n\!}$ by \eqref{RL:reward}
        \STATE Store $\left(s_{n}, a_{n}, r_{n}, s_{n+1}\right)$ into $\mathcal{D}$
        \STATE Replace the oldest tuple if $\left|\mathcal{D}\right| \geq l_{\mathcal{D}}$
        \STATE Sample a random mini-batch $(s,a,r,s')\in \mathcal{B}$
        \STATE Update $\theta$ with the gradient descent method in \eqref{RL:Loss}
        % \STATE Do a gradient descent step according to \eqref{RL:Loss}
        % \STATE Replace target parameter ${\theta}^{-}\leftarrow\theta$ every $l^{-}$ steps 
        \IF{ $G_{n+1} > G_{max}$ } 
            \STATE Update $G_{max} \leftarrow G_{n+1}$
            \STATE Set $s_{n+1} \leftarrow s_0$
        \ELSIF{$G_{n+1}$ \textless $\bar{G}/2$}
            \STATE Set $s_{n+1} \leftarrow s_0$
        \ENDIF
        \STATE Update $n \leftarrow n+1$ and $\theta^- \leftarrow \theta$
    \UNTIL{convergence} 
    % \ENDFOR
\end{algorithmic}
\end{algorithm}

% \vspace*{-3mm}
\section{Numerical Results and Discussions} \label{sec:numerical results}
% \vspace*{-1mm}

In this section, we compare the performance of the proposed codebook with conventional schemes. Table \ref{simul_SetUp} lists the detailed simulation configurations. To conduct practical simulations, all the experiments are based on cross-polarized antenna systems.
We name the proposed LP codebook with the 3GPP based bit allocation introduced in Section \RNum{5} as ``\emph{3GPP based LP}". Also, the LP codebook scheme with the optimal bit allocation scheme given by Algorithm 1 in Section \RNum{5} is named as ``\emph{RL based LP}".

\begin{table}
    \centering
    \caption{Simulation Setup}
    % \vspace*{-3mm}
    \begin{tabular}[ht!]{|l|c|}
        \hline
        Beamforming  &  zero-forcing (ZF)\\
        \hline
        1-to-$M$ mapping $R_{\mathrm{v}}$  &  8 \\ \cline{1-2}
        \hline
        Number of over-sampled DFT beams $L$  &  2 \\ \cline{1-2}
        \hline
        Number of users $K$  &  3\\ \cline{1-2}
        \hline
        Ground distance between a BS and a user $d_{k}$  &  100 m \\ \cline{1-2}
        \hline
        Carrier frequency  &  900 MHz \\ \cline{1-2}
        \hline
        Path-loss $\gamma\left(r_{k}\right)$  &  $8+37.6\log  r_{k} $ \\ \cline{1-2}
        \hline
        \tabincell{l}{Antenna distance $Y_{N}$, $Z_{N}$}  &  0.7$\lambda$, 0.5$\lambda$ \\ \cline{1-2}
        \hline
        Height of a BS and a UE  &  30 m, 2 m \\ \cline{1-2}
        \hline
        Angle-of-arrival in azimuth domain ${\theta}_{k}^{s}$  &  $\left[0,\pi\right]$ \\ \cline{1-2}
        \hline
        Angle-of-arrival in elevation domain ${\phi}_{k}^{s}$  &  $\left[0,\frac{\pi}{36}\right]$ \\ \cline{1-2}
        \hline
        Number of paths $S$  &  20 \\ \cline{1-2}
        \hline
        Transmit power at a BS, $P_{\mathrm{t}}$  &  10 dBm \\ \cline{1-2}
        \hline
        Bandwidth  &  4 MHz \\ \cline{1-2}
        \hline
        Noise figure  &  3 dB \\ \cline{1-2}
        \hline
        Amplification factor for reward calculation $\eta$  &  1000  \\ \cline{1-2}
        \hline
        % Number of episodes $E$  &  1000  \\ \cline{1-2}
        % \hline
        Mini-batch size $|\mathcal{B}|$  &  128  \\ \cline{1-2}
        \hline
        Experience-replay memory size $l_{\mathcal{D}}$  &  2000  \\ \cline{1-2}
        \hline
        % Initial and minimum epsilon ${\epsilon}_{0}$, ${\epsilon}_{min}$  &  1, 0.01  \\ \cline{1-2}
        % \hline
        % Rate of the epsilon decay $r_{\epsilon}$  &  0.99  \\ \cline{1-2}
        % \hline
        Reward discount factor $\gamma$  &  0.99  \\ \cline{1-2}
        \hline
        Learning rate $\alpha$  &  0.001  \\ \cline{1-2}
        \hline
    \end{tabular}
    \label{simul_SetUp}
\end{table}

\begin{figure}
    \begin{center}
    \includegraphics[width=3.6in]{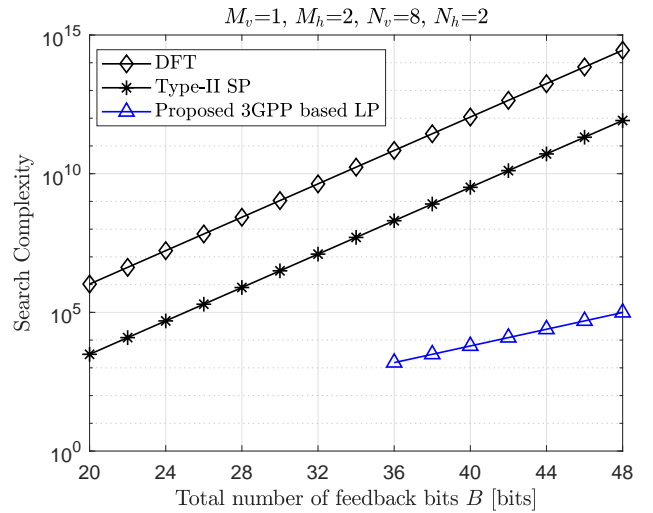}  % 2column
    \end{center}
    \vspace{-2mm}
    \caption{Search Complexity with respect to the total number of feedback bits $B$}
    \label{SearchComplexity}
    \vspace*{-2mm}
\end{figure}

Fig. \ref{SearchComplexity} provides the search complexity of the conventional 2D DFT codebook, the Type-\RNum{2} SP codebook and the proposed 3GPP based LP codebook with respect to the number of feedback bits $B$. The antenna ports $N_{\mathrm{v}}^{\mathrm{t}}$ and $N_{\mathrm{h}}^{\mathrm{t}}$ in \eqref{CBLength_type2} correspond to $M_{\mathrm{v}} N_{\mathrm{v}} / 8 = 1$ and $M_{\mathrm{h}} N_{\mathrm{h}} = 4$, respectively. For the LP scheme, \eqref{CBLength_LPC} leads to $M_{\mathrm{LP}} = M_{\mathrm{h}} = 2$ and $N_{\mathrm{LP}} = M_{\mathrm{v}}N_{\mathrm{v}} N_{\mathrm{h}} / 8 = 2$. 
The required minimum bits on each component are $B_{\mathrm{LP}}=2, ~B_{\mathrm{v}}=0, ~B_{\mathrm{h}}=0, ~B_{\mathrm{p}}=3, ~B_{\mathrm{c}}=2$, and the minimum $B_{\mathrm{L}}$ is 36 in \eqref{Bit_LPC}.
We notice here that although the required minimum bit $B$ for the 3GPP based LP codebook is as large as 36 bits, the advantage of the proposed LP codebook in terms of search complexity is significant. For instance, the search complexity of the 3GPP based LP codebook when $B=48$ is close to that of the Type-\RNum{2} SP codebook with $B=24$.

\begin{figure}
    \begin{center}
    \includegraphics[width=3.6in]{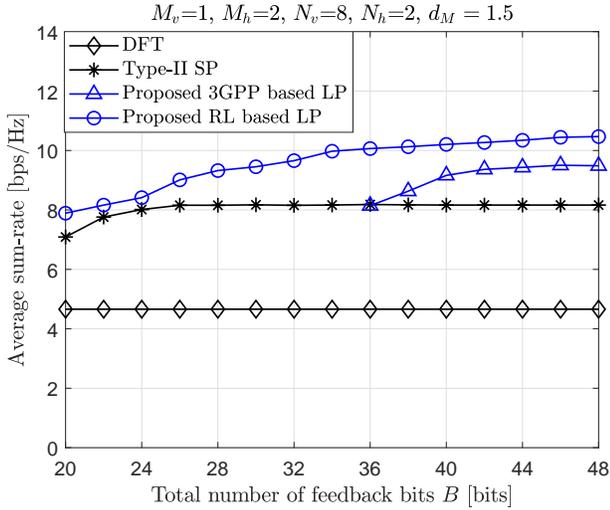}
    \end{center}
    \vspace{-2mm}
    \caption{Average sum-rate with respect to the number of feedback bits $B$}
    \label{XPol_1toM_DQN_M1x2_B}
    \vspace*{-2mm}
\end{figure}

Fig. \ref{XPol_1toM_DQN_M1x2_B} exhibits the average sum-rate in terms of the number of feedback bits $B$. In the figure, the DFT codebook shows no performance change over $B$. A theoretical proof of the inadequacy of the conventional DFT was given in \cite{FYuan:11} for MU-MIMO systems. According to \cite{FYuan:11}, when the angle spread between antennas is large, or the size of the DFT codebook is large, the sum rate of the MU-MIMO system does not improve with the feedback bits. This explains why the DFT scheme in our numerical results does not exhibits noticeable gains. 
The plot also shows that the proposed RL based bit allocation outperforms other schemes for all $B$. Especially when $B=40$, a performance gain of the proposed LP based on RL bit allocation over the Type-\RNum{2} SP scheme reaches $24.8\%$.

\begin{figure}
    \begin{center}
    \includegraphics[width=3.6in]{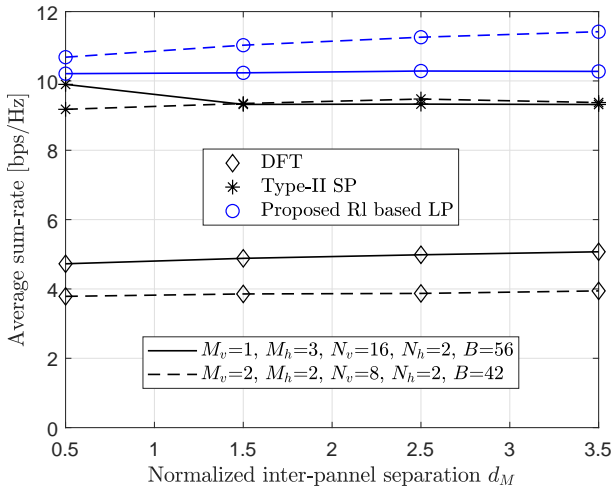} 
    \end{center}
    \vspace{-2mm}
    \caption{Average sum-rate with respect to the normalized panel distance $d_{\mathrm{M}}$}
    \label{XPol_1toM_DQN_dM}
    \vspace*{-2mm}
\end{figure}

Fig. \ref{XPol_1toM_DQN_dM} presents the average sum-rate performance with respect to the normalized panel distance $d_{\mathrm{M}}$. Here, the bit allocation obtained from the RL approach is adopted. First, it can be observed that the RL based LP codebook outperforms the Type-\RNum{2} SP codebook, since the LP codebook exploits panel co-phasing information. The average sum-rate performance gap between these two codebook designs is narrow when $d_{\mathrm{M}}$ equals the antenna distance. Since the PA issue becomes distinct as more panels are combined in the system, the performance of the RL based LP codebook improves with $d_{\mathrm{M}}$.

% \vspace{-2mm}
\section{Conclusion}
\label{sec:conclusion}
% \vspace{-1mm}

In summary, our proposed LP codebook scheme compensates the PA issue by the panel co-phasing factors and individually quantizes the CDI of each line-panel. Moreover, the LP codebook inherits a high degree of freedom from the Type-II SP codebook. These features make the proposed LP codebook achieve the highest gain among all codebook schemes. For the Type-\RNum{2} SP codebook, neglecting the discontinuity between different panels results in a performance loss. Also, we confirm that the LP codebook with RL-based bit allocation designs is an efficient and low-complexity scheme. Furthermore, MPA can be considered as a promising technique for massive MIMO and mmWave systems, and our future work may explore hybrid precoding designs for MPA systems. Reducing feedback bits and feedback delay can also be an interesting future work.

\bibliographystyle{ieeetr}
\begingroup
\renewcommand{\baselinestretch}{0.94}
\bibliography{AZREF}
\endgroup
\end{document}